\def\mpl{m_{\cal P}}
\documentstyle[twoside,12pt]{article}

\catcode`\@=11
\long\def\@makefntext#1{ 
\protect\noindent \hbox to 3.2pt {\hskip-.9pt
{{\twelverm\@thefnmark}}$\hfil}#1\hfill} 

\def\thefootnote{\fnsymbol{footnote}}
\def\@makefnmark{\hbox to 0pt{$^{\@thefnmark}$\hss}}  

\def\ps@myheadings{\let\@mkboth\@gobbletwo
\def\@oddhead{\hbox{} 
\rightmark\hfil\twelverm\thepage}
\def\@oddfoot{}\def\@evenhead{\twelverm\thepage\hfil 
\leftmark\hbox{}}\def\@evenfoot{}
\def\sectionmark##1{}\def\subsectionmark##1{}}

\oddsidemargin=\evensidemargin
\newcommand{\symbolfootnote}{\renewcommand{\thefootnote}
	{\fnsymbol{footnote}}}
\renewcommand{\thefootnote}{\fnsymbol{footnote}}
\newcommand{\alphfootnote}
	{\setcounter{footnote}{0}
	 \renewcommand{\thefootnote}{\sevenrm\alph{footnote}}}

\newcounter{sectionc}\newcounter{subsectionc}\newcounter{subsubsectionc}
\renewcommand{\section}[1] {\vspace{14pt}\addtocounter{sectionc}{1}
\setcounter{subsectionc}{0}\setcounter{subsubsectionc}{0}\noindent
	{\twelvebf\thesectionc. #1}\par\vspace{7pt}}
\renewcommand{\subsection}[1] {\vspace{14pt}\addtocounter{subsectionc}{1}
	\setcounter{subsubsectionc}{0}\noinndent
	{\bf\thesectionc.\thesubsectionc. {\kern1pt \bfit #1}}\par\vspace{7pt}}
\renewcommand{\subsubsection}[1]
{\vspace{14pt}\addtocounter{subsubsectionc}{1}
	\noindent{\twelverm\thesectionc.\thesubsectionc.\thesubsubsectionc.
	{\kern1pt \twelveit #1}}\par\vspace{7pt}}
\newcommand{\nonumsection}[1] {\vspace{14pt}\noindent{\twelvebf #1}
	\par\vspace{7pt}}

\newcounter{appendixc}
\newcounter{subappendixc}[appendixc]
\newcounter{subsubappendixc}[subappendixc]
\renewcommand{\thesubappendixc}{\Alph{appendixc}.\arabic{subappendixc}}
\renewcommand{\thesubsubappendixc}
	{\Alph{appendixc}.\arabic{subappendixc}.\arabic{subsubappendixc}}

\renewcommand{\appendix}[1] {\vspace{12pt}
        \refstepcounter{appendixc}
        \setcounter{figure}{0}
        \setcounter{table}{0}
        \setcounter{lemma}{0}
        \setcounter{theorem}{0}
        \setcounter{corollary}{0}
        \setcounter{definition}{0}
        \setcounter{equation}{0}
        \renewcommand{\thefigure}{\Alph{appendixc}.\arabic{figure}}
        \renewcommand{\thetable}{\Alph{appendixc}.\arabic{table}}
        \renewcommand{\theappendixc}{\Alph{appendixc}}
        \renewcommand{\thelemma}{\Alph{appendixc}.\arabic{lemma}}
        \renewcommand{\thetheorem}{\Alph{appendixc}.\arabic{theorem}}
        \renewcommand{\thedefinition}{\Alph{appendixc}.\arabic{definition}}
        \renewcommand{\thecorollary}{\Alph{appendixc}.\arabic{corollary}}
        \renewcommand{\theequation}{\Alph{appendixc}.\arabic{equation}}
        \noindent{\twelvebf Appendix \theappendixc #1}\par\vspace{5pt}}
\newcommand{\subappendix}[1] {\vspace{12pt}
        \refstepcounter{subappendixc}
        \noindent{\bf Appendix \thesubappendixc. {\kern1pt \bfit #1}}
	\par\vspace{5pt}}
\newcommand{\subsubappendix}[1] {\vspace{12pt}
        \refstepcounter{subsubappendixc}
        \noindent{\rm Appendix \thesubsubappendixc. {\kern1pt \twelveit #1}}
	\par\vspace{5pt}}

\newcommand{\textlineskip}{\baselineskip=14pt}
\newcommand{\smalllineskip}{\baselineskip=12pt}

\def\eightcirc{
\begin{picture}(0,0)
\put(4.4,1.8){\circle{6.5}}
\end{picture}}
\def\eightcopyright{\eightcirc\kern2.7pt\hbox{\eightrm c}}

\newcommand{\copyrightheading}[1]
	{\vspace*{-2.5cm}\smalllineskip{\flushleft
	{\twelverm International Journal of Modern Physics D, #1}\\
	{\twelverm $\eightcopyright$\, World Scientific Publishing
	 Company}\\
	 }}

\def\abstracts#1#2#3{{
	\centerline{ABSTRACT}
	\parindent=0pt #1\par
	\parindent=15pt #2\par
	\parindent=15pt #3
\par}}

\def\keywords#1{{
	\centering{\begin{minipage}{4.5in}\baselineskip=14pt\tenrm
	{\twelveit Keywords}\/: #1
	 \end{minipage} }\par }}

\newcommand{\bibit}{\nineit}
\newcommand{\bibbf}{\ninebf}
\renewenvironment{thebibliography}[1]			
	{\twelverm
	 \baselineskip=14pt				
	 \begin{list}{\arabic{enumi}.}
	{\usecounter{enumi}\setlength{\parsep}{0pt}
	 \setlength{\leftmargin 17pt}{\rightmargin 0pt}	
	 \setlength{\itemsep}{0pt} \settowidth		
	{\labelwidth}{#1.}\sloppy}}{\end{list}}

\newcounter{itemlistc}
\newcounter{romanlistc}
\newcounter{alphlistc}
\newcounter{arabiclistc}

\newcommand{\fcaption}[1]{
        \refstepcounter{figure}
        \setbox\@tempboxa = \hbox{\tenrm Fig.~\thefigure. #1}
        \ifdim \wd\@tempboxa > 5in
           {\begin{center}
        \parbox{5in}{\tenrm \smalllineskip Fig.~\thefigure. #1 }
            \end{center}}
        \else
             {\begin{center}
             {\tenrm Fig.~\thefigure. #1}
              \end{center}}
        \fi}

\newcommand{\tcaption}[1]{
        \refstepcounter{table}
        \setbox\@tempboxa = \hbox{\tenrm Table~\thetable. #1}
        \ifdim \wd\@tempboxa > 5in
           {\begin{center}
        \parbox{5in}{\tenrm\smalllineskip Table~\thetable. #1 }
            \end{center}}
        \else
             {\begin{center}
             {\tenrm Table~\thetable. #1}
              \end{center}}
        \fi}

\def\@citex[#1]#2{\if@filesw\immediate\write\@auxout	
	{\string\citation{#2}}\fi			
\def\@citea{}\@cite{\@for\@citeb:=#2\do			
	{\@citea\def\@citea{,}\@ifundefined		
	{b@\@citeb}{{\bf ?}\@warning
	{Citation `\@citeb' on page \thepage \space undefined}}
	{\csname b@\@citeb\endcsname}}}{#1}}

\newif\if@cghi
\def\cite{\@cghitrue\@ifnextchar [{\@tempswatrue
	\@citex}{\@tempswafalse\@citex[]}}
\def\citelow{\@cghifalse\@ifnextchar [{\@tempswatrue
	\@citex}{\@tempswafalse\@citex[]}}
\def\@cite#1#2{{$\null^{#1}$\if@tempswa\typeout
	{IJCGA warning: optional citation argument
	ignored: `#2'} \fi}}
\newcommand{\citeup}{\cite}

\def\pmb#1{\setbox0=\hbox{#1}
	\kern-.025em\copy0\kern-\wd0
	\kern.05em\copy0\kern-\wd0
	\kern-.025em\raise.0433em\box0}
\def\mbi#1{{\pmb{\mbox{\scriptsize ${#1}$}}}}
\def\mbr#1{{\pmb{\mbox{\scriptsize{#1}}}}}

\def\fnm#1{$^{\mbox{\scriptsize #1}}$}
\def\fnt#1#2{\footnotetext{\kern-.3em
	{$^{\mbox{\scriptsize #1}}$}{#2}}}

\def\fpage#1{\begingroup
\voffset=.3in
\thispagestyle{empty}\begin{table}[b]\centerline{\footnotesize #1}
	\end{table}\endgroup}

\def\runninghead#1#2{\pagestyle{myheadings}
\markboth{{\tenit{\quad #1}}\hfill}{\hfill{\tenit{#2\quad}}}}

\font\it=cmti12
\font\bf=cmbx12
\font\twelvebf=cmbx12
\font\twelveit=cmti12
\font\twelveit=cmti12
\font\bfit=cmbxti10 at 10pt
\font\twelvebf=cmbx12 
\font\twelverm=cmr12  
\font\twelveit=cmti12 
\font\tenbf=cmbx10
\font\tenrm=cmr10
\font\tenit=cmti10
\font\tenit=cmti10
\font\ninerm=cmr9
\font\sevenrm=cmr7

\newcommand{\proof}[1]{{\tenbf Proof.} #1 $\Box$.}

\def\qed{\hbox{${\vcenter{\vbox{                          
   \hrule height 0.4pt\hbox{\vrule width 0.4pt height 6pt
   \kern5pt\vrule width 0.4pt}\hrule height 0.4pt}}}$}}

\begin{document}
\normalsize\textlineskip
{\thispagestyle{empty}
\setcounter{page}{1}
\pagestyle{empty}

\renewcommand{\thefootnote}{\fnsymbol{footnote}} 

%
%

\centerline{\hfill ALBERTA THY/24-94}
\vspace{0.5truein}
\centerline{\bf RESOLVING THE QUESTION OF TIME FOR}
\vspace{0.3truein}
\centerline{\bf SEMI-CLASSICAL GRAVITY }
\vspace{1.0truein}
\centerline{\bf D.S. SALOPEK}
\vspace{0.4truein}
\centerline{\it Department of Physics}
\centerline{\it University of Alberta}
\centerline{\it Edmonton, Canada T6G 2J1 }
%
\vspace{1.0truein}
\abstracts{
\noindent
Hamilton-Jacobi theory provides a natural
starting point for a covariant description of the gravitational field.
Using a spatial gradient expansion, one may solve
for the phase of the wavefunction by using a line-integral
in superspace. Each contour of integration corresponds to a particular
choice of the time-hypersurface, and each yields
the same answer. In this way, one can describe all time
choices simultaneously. As a demonstration of the formalism, I
will compute
large-angle microwave background anisotropies and the
galaxy-galaxy correlation function associated with scalar
and tensor fluctuations of power-law inflation. }{}{}
\vspace{1.0truein}
\centerline{In Proceedings of}
\centerline{Lake Louise Winter Institute}
\centerline{\it Particle Physics and Cosmology}
\centerline{February 20-26, 1994}
\centerline{Eds. B. Campbell and F. Khanna, World Scientific}
\vfill\eject
\vspace*{-3pt}\textlineskip
\section{Introduction}
\noindent
Hamilton-Jacobi (HJ) theory is a cornerstone of modern
theoretical physics. It may be profitably applied
to numerous problems in cosmology, and it is particularly
useful in describing the inflationary scenario.
Since a full quantum theory of the gravitational field is
lacking, we should be content with a semi-classical treatment
(at least for now).

HJ theory has been successfully employed in deriving the Zel'dovich
approximation\cite{Zel} (which describes the formation
of sheet-like structures in the Universe) and its higher
order generalizations from
general relativity.\cite{CPSS94}{}$^{-}$\cite {SSC94}
Various researchers have
employed HJ methods in an attempt to recover the inflaton potential
from cosmological observations.\cite{COPELAND93} Moreover,
they can be used to construct inflationary models that
yield non-Gaussian primordial fluctuations;\cite{SB1} such models
could possibly resolve the problem of large scale structure.\cite{Mosc93}

Here I will focus on one particularly attractive feature of
HJ theory: it provides a covariant formulation of the
gravitational field. In the semi-classical theory, the answer
to the question of time is clear: time is arbitrary.
What is required then is a formalism that
treats all time choices simultaneously, and
I will demonstrate that HJ theory passes the test.
Before describing the general technique, I will
consider a simple analogy from potential theory
which illuminates the essential point.

\section{Potential Theory}

The fundamental problem in potential theory is: given a force
field $g^i(u_k)$ which is a function of $n$ variables $u_k$,
what is the potential $\Phi \equiv \Phi(u_k)$ (if it exists)
whose gradient returns the force field,
\begin{equation}
{\partial \Phi \over \partial u_i} = g^i(u_k) \quad ?
\end{equation}
Not all force fields are derivable from
a potential. Provided that the force field satisfies the
integrability relation,
\begin{equation}
0= {\partial g^i \over \partial u_j} - {\partial g^j \over \partial u_i} =
\left [{\partial  \over \partial u_j}, {\partial  \over  \partial u_i }
\right ] \, \Phi \, ,
\end{equation}
(i.e., it is curl-free),
one may find a solution which is conveniently expressed using a
line-integral
\begin{equation}
\Phi(u_k) = \int_C \sum_j dv_j \ g^j(v_l) \ .
\end{equation}
If the two endpoints are fixed, all contours return the same
answer. In practice, I will employ the simplest contour that
one can imagine: a line connecting the origin to the
observation point $u_k$. Using $s$, $0 \le s \le 1$,
to parameterize the contour, the line-integral may be rewritten as
\begin{equation}
\Phi(u_k) = \sum_{j=1}^n \int_0^1  ds  \; u_j \ g^j(su_k) \ .
\label{lint}
\end{equation}
Similarly, in solving for the phase of the wavefunctional, I
will employ a line-integral in {\it superspace}.

\section{Solving the Hamilton-Jacobi Equation for General Relativity}
\noindent
The Hamilton-Jacobi equation for general relativity is
derived using a Hamiltonian formulation of gravity.
One first writes the line element using the ADM 3+1 split,
\begin{equation}
ds^2=\left(-N^2+\gamma^{ij}N_iN_j\right)dt^2 + 2N_idt\,dx^i +
\gamma_{ij}dx^i\,\
dx^j\ ,
\label{ADMdecomp}
\end{equation}
where $N$ and $N_i$ are the lapse and shift functions, respectively,
and $\gamma_{ij}$ is the 3-metric. Hilbert's action for gravity interacting
with a scalar field becomes
\begin{equation}
{\cal I}=\int d^4x\left(\pi^{\phi}\dot\phi +\pi^{ij}\dot\gamma_{ij}
-N{\cal H} -N^i{\cal H}_i\right).
\label{ADMaction}
\end{equation}
The lapse and shift functions are Lagrange multipliers that
ensure that the energy constraint ${\cal H}(x)$ and the
momentum constraint ${\cal H}_i(x)$ vanish.

The object of chief importance is the generating functional
${\cal S}\equiv {\cal S}[\gamma_{ij}(x), \phi(x)]$.
For each universe with field configuration
$[\gamma_{ij}(x), \phi(x)]$ it assigns a number
which can be complex. The generating functional is
the `phase' of the wavefunctional in the semi-classical approximation:
$\Psi \sim e^{i{\cal S}}$.
For the applications that we are considering, the prefactor
before the exponential is not very important, although
it has interesting consequences for quantum
cosmology.\cite{B93} The probability functional,
${\cal P} \equiv |\Psi|^2$, is given by the square of the wavefunctional.

Replacing the conjugate momenta by functional derivatives
of ${\cal S}$ with respect to the fields,
\begin{equation}
\pi^{ij}(x)={\delta{\cal S}\over \delta{\gamma_{ij}(x)}}\ , \qquad
\pi^{\phi}(x)={\delta{\cal S}\over \delta\phi (x)}\ ,
\label{pis}
\end{equation}
and substituting into the energy constraint,
one obtains the Hamilton-Jacobi equation,
\begin{eqnarray}
{\cal H}(x)=&&\gamma^{-1/2} {\delta{\cal S}\over \delta\gamma_{ij}(x)}
{\delta{\cal S}\over \delta\gamma_{kl}(x)}
\left[2\gamma_{il}(x) \gamma_{jk}(x) - \gamma_{ij}(x)\gamma_{kl}(x)\right]
\nonumber \\
&& + {1\over 2} \gamma^{-1/2}
\left({\delta{\cal S}\over \delta\phi(x)}\right)^2
+\gamma^{1/2}V(\phi(x)) \nonumber \\
&& -{1\over 2}\gamma^{1/2}R
+{1\over 2} \gamma^{1/2}\gamma^{ij}\phi_{,i}\phi_{,j}=0 \ ,
\label{HJequation}
\end{eqnarray}
which describes how ${\cal S}$ evolves in superspace.
$R$ is the Ricci scalar associated with the 3-metric, and $V(\phi)$
is the scalar field potential.
In addition, one must also satisfy the momentum constraint
\begin{equation}
{\cal H}_{i}(x)=-2\left(\gamma_{ik}{\delta{\cal S}\over \delta\gamma_{kj}(x)}
\right)_{,j} +
{\delta{\cal S}\over\delta\gamma_{lk}(x)}\gamma_{lk,i} +
{\delta{\cal S}\over\delta\phi (x)} \phi_{,i}=0 \ ,
\label{Smomentum}
\end{equation}
which legislates that ${\cal S}$ be invariant under
reparametrizations of the spatial coordinates.
(Units are chosen so that
$c=8\pi G= \hbar= 1$). Since neither the lapse function
nor the shift function appears in
eqs.(\ref{HJequation},\ref{Smomentum}) the temporal
and spatial coordinates are {\it arbitrary}:
HJ theory is {\it covariant}.

In order to solve eqs.(\ref{HJequation},\ref{Smomentum}),
I will expand the generating functional
\begin{equation}
{\cal S}= {\cal S}^{(0)} + {\cal S}^{(2)} + {\cal S}^{(4)} + \dots\ ,
\label{theexpansion}
\end{equation}
in a series of terms according to
the number of spatial gradients that they contain.
As a result, the Hamilton-Jacobi equation can likewise be
grouped into terms with an even number of spatial
derivatives:
\begin{equation}
{\cal H}={\cal H}^{(0)} + {\cal H}^{(2)} + {\cal H}^{(4)} + \dots\ .
\label{theexpansion2}
\end{equation}
The invariance of the generating functional
under spatial coordinate transformations
suggests a solution of the form,
\begin{equation}
{\cal S}^{(0)}[ \gamma_{ij}(x), \phi(x)] = - 2 \int d^3x
\gamma^{1/2} H \left[ \phi(x) \right] \ ,
\end{equation}
for the zeroth order term ${\cal S}^{ (0) }$. The function
$H \equiv H(\phi)$ satisfies the separated HJ equation of order zero,
\begin{equation}
H^2={2\over 3}\left( {\partial H\over\partial\phi} \right)^2
 +{1 \over 3}V\left( \phi \right ) \ ,
\label{Hequation}
\end{equation}
which is an ordinary differential equation.
Note that ${\cal S}^{(0)}$ contains no spatial gradients.

In order to compute the higher order terms, one introduces a
change of variables, $( \gamma_{ij}, \phi ) \rightarrow
(f_{ij}, u)$:
\begin{equation}
u = \int \  { d \phi \over -2 { \partial H \over \partial \phi } } \ ,
\quad f_{ij} = \Omega^{-2}(u) \, \gamma_{ij} \ , \label{changeA}
\end{equation}
where the conformal factor $\Omega \equiv \Omega(u)$ is defined through
\begin{equation}
{ d \ln \Omega \over d u} \equiv -2 { \partial H \over \partial \phi}
{ \partial \ln \Omega \over \partial \phi} = H \ . \label{changeB}
\end{equation}
in which case the equation for ${\cal S}^{(2m)}$ becomes
\begin{equation}
{\delta {\cal S}^{(2m)}\over\delta u(x)}\Bigg|_{f_{ij}}
+ {\cal R}^{(2m) }[u(x), f_{ij}(x)] =0\ .
\label{HJ.conf.for.dust}
\end{equation}
The remainder term ${\cal R}^{(2m) }$ depends on some quadratic
combination of the previous order terms (it may be written explicitly).
For example, for $m=1$, it is
\begin{equation}
{\cal R}^{(2)} =
{1\over 2} \gamma^{1/2}\gamma^{ij}\phi_{,i}\phi_{,j}
-{1\over 2}\gamma^{1/2}R \, .
\end{equation}
Eq.(\ref{HJ.conf.for.dust}) has the form of an infinite dimensional gradient.
It may integrated using a line integral analogous to
eq.(\ref{lint}):
\begin{equation}
{\cal S}^{(2m)}=-\int d^3x\int_0^1  ds\  u(x) \
{\cal R}^{(2m)}[su(x), f_{ij}(x)] \; ;
\label{lints}
\end{equation}
the conformal 3-metric $f_{ij}(x)$ is held constant
during the integration which may be performed explicitly
in many cases of interest.

The integrability
condition for the HJ equation\cite{MT72} follows from the
Poisson bracket of the energy constraints evaluated at
spatial points $x$ and $x^\prime$,
\begin{equation}
\{ {\cal H}(x^k), {\cal H}(x^{k^\prime}) \} =
[ \gamma^{ij}(x^k) {\cal H}_j(x^k)+
\gamma^{ij}(x^{k^\prime}) {\cal H}_j(x^{k^\prime}) ]
\; \delta^3_{,i}(x^k - x^{k^\prime}) \, .
\label{poisson}
\end{equation}
In fact, alternative contours describing the
line-integral (\ref{lints}) will
correspond to different time-hypersurface choices.
Provided that the generating functional is invariant
under reparametrizations of the spatial coordinates,
(e.g., ${\cal H}_i$ vanishes in the right-hand-side of
eq.(\ref{poisson})),
different time-hypersurface choices will lead to the same
generating functional. Hypersurface invariance is closely
related to gauge-invariance. Hence the line-integral solution goes
a long way in understanding the role of time in semi-classical
gravity.
\vfill\eject
\section{Computing Large-Angle Microwave Background Fluctuations
and Galaxy Correlations}
\noindent
In order to describe the fluctuations arising during the
inflationary epoch, it is necessary to sum an infinite
subset\cite{SS94} of the terms ${\cal S}^{(2m)}$.
In this case, one considers all terms which are quadratic in the Ricci
tensor $\tilde R_{ij}$ of the conformal 3-metric $f_{ij}(x)$.
Once again, no explicit choice of time hypersurface is made.

However, when one compares with observations, there
are indeed preferred gauges. The phase transition
where photons decouple from matter
occurs essentially on a uniform temperature slice,
$T \sim 4000 K$, when protons combine with electrons
to form neutral hydrogen.  For adiabatic
perturbations at large wavelengths, this
is the same as a comoving, synchronous time slice
which was the choice made by Sachs and Wolfe\cite{SW67}
in the computation of large-angle temperature anisotropies.

Here I will be content to graphically display the
final results for the power-law inflationary model
where the scalar field potential is given by
\begin{equation}
V(\phi)= V_0 \; {\rm exp} \left(-\sqrt{2\over p} \phi \right)
\label{s.f.potential}
\end{equation}
where $p$ is a constant that determines the steepness of the potential.
This model is of high interest because it may produce
copious amounts of primordial gravitational radiation, which
is in essence a quantum gravitational effect.

For various values of $p$,
Fig.(1) shows the power spectrum, ${\cal P}_\zeta(k)$,
for $\zeta$ which parametrizes the primordial scalar perturbations
associated with the comoving wavenumber $k$.
(The present value of the Hubble parameter is assumed
to be $H_0=$ $50$ $km \; s^{-1}\;  Mpc^{-1}$.)
The power spectra have been normalized using the 2-year data
set\cite{BENNETT94} of the DMR (Differential Microwave Radiometer) experiment
on board of the COBE (Cosmic Background Explorer) satellite:
\begin{equation}
\sigma_{sky}(10^0)= 30.5 \pm 2.7 \mu K \quad (68 \; \% \;
{\rm confidence \;level}).
\end{equation}
The spectral index $n_s$ for scalar perturbations
is defined by
\begin{equation}
n_s \equiv 1+ { d \; {\rm log}_{10} {\cal P}_\zeta(k) \over
d \; {\rm log}_{10} (k) } =
1- 2/(p-1)  \, .
\end{equation}
As $p \rightarrow \infty$, one recovers the flat
Zel'dovich spectrum $n_s=1$, and there is no graviton
production. As $p$ decreases, the spectrum
tilts giving more power at larger scales, and the graviton
production increases, contributing $50 \%$ to COBE's signal
at $n_s=0.8$. (For comparison, if there were no graviton
production, essentially all the curves would join at $k=10^{-4}$Mpc
which is the length scale effectively probed by COBE).

In Fig.(2), I plot the linear density contrast at the present
epoch for the same set of models, assuming the cold-dark-matter
model. The bold line shows the observed clustering of
galaxies: $\xi_{gg}(r)= (r/ r_0)^{-\gamma}$  where $r_0= 10$ Mpc,
$\gamma=1.8$. In order that there
be enough fluctuations to seed galaxies, one requires
that the biasing parameter $b_\rho$ be less than 2 which implies that
\begin{equation}
n_s> 0.8 \, , \quad {\rm (power-law \; inflation)}.
\label{plil}
\end{equation}
This constraint was
more stringent than the lower limit given by COBE's first
year data set: $n_s= 1.1^{+0.45}_{-0.65}$
($68 \%$ confidence level).

\begin{figure}[htbp]
\setlength{\unitlength}{0.240900pt}
\ifx\plotpoint\undefined\newsavebox{\plotpoint}\fi
\sbox{\plotpoint}{\rule[-0.175pt]{0.350pt}{0.350pt}}%

\fcaption{ Primordial scalar perturbations of the metric
are described by the function $\zeta$. The power spectra for
zeta are shown
for various choices of the the spectral index $n_s= 1- 2/(p-1)$
arising from power-law inflation. They have been normalized using
COBE's 2-yr data set. }
\end{figure}

Using two years of data, the COBE DMR team\cite{GORSKI94}
now reports $n_s=1.10 \pm 0.32$
($68 \%$ confidence level). Here they have included
the quadrupole amplitude in their computations.
However in determining the quadrupole, one must
subtract out the contribution of the Milky Way
which is quite tricky to do in practice. For this reason, one
may wish to exclude the quadrupole component in which case they obtain
$n_s=0.87  \pm 0.36$. Either way,
their results are still consistent with the simplest
models of inflation which yield $n_s < 1$.
In addition, they are consistent with the limit for
power-law inflation eq.(\ref{plil}).

\section{Conclusions}
\noindent
The question of the choice of time is an extremely difficult one,
particularly for the quantum theory of the gravitational field.
Here I have shed some light on the semi-classical problem.
The choice of time is arbitrary, and using Hamilton-Jacobi
theory one may construct a covariant formalism which treats
all time choices on an equal footing. A line-integral
in superspace allows one to solve for the phase of the
wavefunctional. Different contours of integration lead to the same
answer provided gauge-invariance is maintained.

\begin{figure}[htbp]
\setlength{\unitlength}{0.240900pt}
\ifx\plotpoint\undefined\newsavebox{\plotpoint}\fi
\sbox{\plotpoint}{\rule[-0.175pt]{0.350pt}{0.350pt}}%

\fcaption{
For the present epoch, the power spectra
for the linear density perturbation $\delta \rho/ \rho$
in comoving synchronous
gauge are shown. The dark line depicts the observed two-point correlation
function describing galaxy clustering. If there is no
biasing, $n_s=0.9$ gives a good fit
to the data near $k=10^{-1}Mpc^{-1}$. In order that there
be enough fluctuations to seed galaxies, one
requires that the biasing parameter $b_\rho$ be less than 2
which implies that $n_s > 0.8 $.
As a result for power-law
inflation, at most $\sim 50\%$ of COBE's signal may arise from gravitational
waves.}
\end{figure}

As an application, I briefly summarized computations of
large-angle microwave background anisotropies and the
galaxy-galaxy correlation function for power-law inflation
used in conjunction with the cold-dark-matter model. This inflation model
is particularly interesting because it may produce large
amounts of primordial gravitational waves. As a result, the
spectral index is restricted to  be $n_s > 0.8$
otherwise there are not enough fluctuations to seed galaxies.

\section{Acknowledgments}

\noindent
I thank John Stewart and Joe Parry for a fruitful
collaboration on these topics.
I also like to thank Profs. Bruce Campbell and
Faqir Khanna for organizing an excellent conference.
This work was supported by NSERC of Canada, and a
CITA National Fellowship held in Edmonton.

\nonumsection{References}

\end{document}